\documentstyle[aps,prl,preprint]{revtex}
\begin{document}
\title
{Raman Scattering in a Two--Layer Antiferromagnet}
\author{Dirk K. Morr,$^{1}$ Andrey V. Chubukov,$^{1,2}$ Arno P.
Kampf,$^{3}$ and G. Blumberg$^{4,5}$}
\address{$^{1}$
Department of Physics, University of Wisconsin--Madison,\\
1150 University Avenue, Madison,
 Wisconsin 53706}
\address{$^{2}$
P.L. Kapitza Institute for Physical Problems, Moscow, Russia}
\address{$^{3}$
Institut f\"ur Theoretische Physik, Universit\"at zu K\"oln, Z\"ulpicher Str.
77,\\
 50937 K\"oln, Germany}
\address{$^{4}$
NSF Science and Technology Center for Superconductivity and\\
Department of Physics, University of Illinois at Urbana-Champaign,\\
1110 W. Green Str. Urbana IL 61801}
\address{$^{5}$
Institute of Chemical Physics and Biophysics, R\"avala 10,
Tallinn EE0001, Estonia}
\date{\today}
\maketitle
\begin{abstract}
Two--magnon Raman scattering is a useful tool to verify recent
suggestions concerning the value of the interplanar exchange constant in
antiferromagnetic two--layer systems, such as $YBa_2Cu_3O_{6+x}$.
We present a theory for Raman scattering in a two--layer antiferromagnet. We
study the spectra for the electronic and magnetic excitations across the charge
transfer gap within the
one--band Hubbard model and derive the matrix elements for the Raman scattering
cross section in a diagrammatic formalism. We analyze the effect of the
interlayer exchange coupling $J_2$ for the Raman spectra in $A_{1g}$ and
$B_{1g}$ scattering geometries both in the non--resonant regime (when the
Loudon--Fleury model is valid), and at resonance. We show that within the
Loudon--Fleury approximation, a nonzero $J_2$ gives rise to a finite signal in
$A_{1g}$ scattering geometry. Both, in this approximation and at resonance,
the intensity in the $A_{1g}$ channel
has a peak at {\it small} transferred frequency equal to twice the
 gap in the spin--wave spectrum. We compare our
results with
experiments in $YBa_2Cu_3O_{6.1}$ and $Sr_2CuO_2Cl_2$ compounds and argue that
the large value of $J_2$ suggested in a number of recent studies is
incompatible with Raman experiments in $A_{1g}$ geometry.
\end{abstract}
\pacs{PACS: ,}
\narrowtext
\section{Introduction}
Since the discovery of high--$T_c$ superconductivity \cite{Bednorz} a lot of
work has been done in an attempt to understand the pairing mechanism. Most of
the existing theories consider boson--mediated pairing between electrons in the
same $CuO_2$ plane, but there also exist arguments in favor of the pairing
between electrons in adjacent $CuO_2$ planes \cite{chakandsudbo}. These
arguments are mostly applied to $YBa_2Cu_3O_{6+x}$ ($YBCO$) compounds in which
the unit cell contains pairs of $CuO_2$ planes separated by a charge reservoir.
A strong magnetic coupling between the planes of a bilayer was also suggested
\cite{Altshuler,Millismonien,Lee} as a possible source for an experimentally
observed strong downturn renormalization of the low temperature Pauli
susceptibility of holes in underdoped $YBCO$ compounds (``spin--gap''
phenomenon) as well as for the maximum in the spin--lattice relaxation rate at
temperatures well above $T_c$ \cite{Takigawa}.

An essential input parameter for these theories is the value of the interplane
hopping amplitude or the $Cu-Cu$ superexchange interaction between the two
$CuO_2$ planes. In this paper, we will argue that the Raman scattering
experiments in two--layer compounds allow an estimate for the value of the
interlayer exchange coupling.

The Raman scattering in single--layer parent high--$T_c$ compounds has been
intensively studied over the last few years
\cite{Mer80,Cot86,Sin91,Sin89,Can92,Chu95}. A large number of studies has been
performed to understand the Raman spectra in different scattering geometries.
In $B_{1g}$ geometry (${\bf e}_i=(\hat{x}+\hat{y})/\sqrt{2},\,
{\bf e}_f=(\hat{x}-\hat{y})/\sqrt{2}$ \cite{xy}, where ${\bf e}_{i,f}$ are
polarization unit vectors of the incident and scattered photons) the dominant
feature of the magnetic Raman intensity profile is a peak at about
$3000 cm^{-1}$, which is attributed to a two--magnon scattering process
\cite{Lyo88}. The two--magnon peak has been observed in all parent high--$T_c$
compounds. Besides, the experiments also found a strong Raman signal in the
$A_{1g}$ geometry (${\bf e}_i={\bf e}_f =(\hat{x}+\hat{y})/\sqrt{2}$)\cite{xx}.
The $A_{1g}$ signal has a maximum at about the same frequency as in $B_{1g}$
geometry, but the width of the peak is larger and its intensity is about a
quarter as strong.

A traditional framework for the understanding of Raman experiments is the
Loudon--Fleury theory \cite{Fle68} which describes the interaction of light
with only spin degrees of freedom. This theory explains a peak in $B_{1g}$
geometry but predicts that there should be no scattering in $A_{1g}$ geometry.
Recently, however, it was found \cite{Chu95} that the Loudon--Fleury approach
has to be modified because Raman experiments are mostly performed near
the resonant regime where photon frequencies are close to the charge transfer
gap of the insulating compounds, and one can by no means neglect electronic
degrees of freedom. In this regime, the diagrams which are neglected in the
Loudon--Fleury theory, and which contribute to both $A_{1g}$ and $B_{1g}$
scattering are actually more important than the diagrams included in the
Loudon--Fleury theory.

In this paper, we will study magnetic Raman scattering in two--layer
antiferromagnetic insulators. We will show that in the presence of the
interlayer exchange coupling, $J_2$, the Raman scattering profile in the
$A_{1g}$
scattering geometry acquires {\it qualitatively} new features. In particular, a
nonzero $J_2$ gives rise to a nonzero Raman intensity in the $A_{1g}$
scattering
geometry already in the Loudon--Fleury approximation. Moreover, we will see
that within the Loudon-Fleury theory, there is a very strong enhancement of
$A_{1g}$ intensity at {\it small} transferred frequencies
$\omega_i-\omega_f=\delta\omega^{res}\approx 4(J_1J_2)^{1/2}$, where $\omega_i$
and $\omega_f$ are the frequencies of the incoming and the scattered photon,
respectively, and $J_1$ is the intralayer exchange coupling. At the frequency
shift $\delta\omega^{res}$, the intensity in the $A_{1g}$ channel actually
 turns out to be {\it larger} than the intensity in the $B_{1g}$ channel,
which, as we recall, is nonzero already in the absence of $J_2$. In the
resonant regime, when the incident photon frequency becomes comparable to the
single particle excitation gap of the insulator, there exists substantial
$A_{1g}$ scattering already for a single layer. Nevertheless, we will argue
that even in this situation, in a two-layer system,
there is a measurable change of the Raman
intensity near the frequency shift $\delta\omega^{res}$. We will also consider
the scattering in $B_{1g}$ geometry and will show that in this geometry the
effect of $J_2$ is much weaker than for $A_{1g}$ scattering.

In principle, the presence of the new features in the $A_{1g}$ Raman scattering
allows one to find the value of the interlayer coupling. In reality, however,
in $YBCO$, the frequency shift $\delta\omega^{res}$ is in the region where the
dominant contribution to the Raman intensity comes from phonon rather than
two--magnon scattering. However, we will argue that there are still several
features of the Raman profile in $YBCO$ which are absent in the single--layer
compound $Sr_2CuO_2Cl_2$ and which allow to find an estimate for $J_2$. We
found from our analysis that $J_2$ is likely to be about $0.1 J_1$. More
rigorously, we can place the upper boundary for $J_2$ as $J_2 \sim 0.25 J$.
Neutron
scattering data for the spin wave velocity $c_{sw}$ at half-filling
\cite{Shamoto} yield  $J_1\sim 120meV$. As
$c_{sw}$ is only weakly dependent on $J_2$ for small $J_2/J_1$, this implies
that the probable value is $J_2\sim 12meV$,
and the upper boundary for $J_2$ is
$30 meV$. This is consistent with the estimate $5meV<J_2<20meV$ for $J_2$
extracted from the analysis of NMR data \cite{MMpreprint} on the double--layer
material $Y_2Ba_4Cu_7O_{15}$ \cite{Stern}, but substantially smaller than
$J_2\sim 0.55 J_1$ inferred from infrared transmission and reflection
measurements~\cite{Gruninger95} and also substantially smaller than the
theoretical estimate $J_2=56meV$ by Barriquand and Sawatzky \cite{Sawatzky}.

The paper is organized as follows. We start in Sec. \ref{matrix} with the
Hubbard model at half--filling which has long range antiferromagnetic order in
its ground state, and derive in a diagrammatic formalism  the effective
Loudon--Fleury model for Raman scattering in $A_{1g}$ and $B_{1g}$ geometries
in the non--resonant regime (i.e. assuming that the photon frequencies are
smaller than the Mott--Hubbard gap). In Sec. \ref{raman-int}, we use this model
to compute the Raman intensity in the $A_{1g}$ and $B_{1g}$ channels first
without magnon--magnon interaction, and then by including multiple scattering
between magnons. In Sec. \ref{resonant-case}, we will discuss the resonant
regime. Finally, in Sec. \ref{discussion}, we summarize our results and discuss
them in the context of  experimental data for single--layer $Sr_2CuO_2Cl_2$ and
double--layer $YBa_2Cu_3O_{6.1}$.

\section{The derivation of the Loudon--Fleury Hamiltonian for the non--resonant
case}
\label{matrix}

In this section we derive the effective Loudon--Fleury model for Raman
scattering using a momentum space diagrammatic formalism. This formalism has
recently been applied to derive the Loudon--Fleury Hamiltonian for a
single--layer system~\cite{Chu95}. Technically, the calculations for two--layer
systems are more involved as one has to double the number of fermionic
operators. Conceptually, however, our approach is exactly the same as in
Ref.~\cite{Chu95}, and we therefore refrain from discussing the calculational
steps in full length.

The starting point of our calculations is the simplest one--band Hubbard
Hamiltonian for a two--layer system at half--filling on a square lattice given
by:
\begin{equation}
H=-t\sum_{<i,j>}{(c_{i,\sigma}^\dagger c_{j,\sigma} + d_{i,\sigma}^\dagger
d_{j,\sigma} + h.c.)} - t' \sum_{i} {(c_{i,\sigma}^\dagger d_{i,\sigma}+ h.c.)}
+ U \sum_{i, \alpha}{ n_{i,\alpha,\uparrow} n_{i,\alpha,\downarrow} }
\label{Hubbard}
\end{equation}
where the $c$ and $d$ operators represent the electrons of layer 1 and 2,
respectively, $\alpha=1,2$, $n_{i,1,\sigma}=c_{i,\sigma}^\dagger c_{i
\sigma},~n_{i,2,\sigma}=d_{i,\sigma}^\dagger d_{i,\sigma}$, and $t'$ is the
hopping
amplitude between the planes for which we assume $t'<t$ (see Fig.
\ref{system}). We will also assume that $t/U\ll 1$ and thus perform our
calculations only to leading order in $t/U$. This large $U$ one--band model
with only nearest--neighbor in--plane hopping is indeed a simplification, but
it was argued in \cite{Chu95} that this model already contains the relevant
physics for the analysis of Raman scattering in $YBCO$ compounds. Here we
follow this reasoning and assume that the model of Eq.~(\ref{Hubbard}) is
valid.

The mechanism of two--magnon Raman scattering is straightforward and has been
discussed a number of times in the literature \cite{Marville}: the incoming
photon with frequency $\omega_i$ creates a virtual particle--hole pair, which
then in turn emits two magnons with momenta ${\bf q}$ and $-{\bf q}$ before it
annihilates into an outgoing photon with frequency $\omega_f$. For the
diagrammatic calculation of the required matrix element for Raman scattering we
thus need to compute two types of vertices: the vertices for the interaction
between the electrons and the electromagnetic vector potential of the photons
and the vertices for the interaction between the electrons and the magnons.

The procedure to derive the coupling of light to the electrons was previously
described by Shastry and Shraiman \cite{Sha90}: the photons introduce a slowly
varying vector potential ${\bf A}({\bf r},t)$ in the presence of which the
hopping term in the kinetic energy of the electrons acquires a phase
$\Big(i{e\over\hbar c}\int_i^j{\bf A}({\bf l},t)\cdot d{\bf l}\Big)$. The
Hubbard Hamiltonian is then expanded to leading orders in ${\bf A}$.
One further introduces the staggered magnetization as the spin density
wave (SDW) order parameter and transforms to new fermionic operators which
diagonalize the Hartree--Fock factorized Hamiltonian \cite{Sch89}. To derive
the interaction vertex between fermions and magnons, one has to compute the
transverse spin susceptibilities with and without momentum transfer
${\bf Q}=(\pi,\pi)$, and construct the interaction Hamiltonian which reproduces
{\it all} dynamic spin susceptibilities. This procedure is unique though rather
involved for a two--layer system. We will skip the details and present only our
final results (some of the useful formulae are collected in Appendix
\ref{en-disp-app}).

In terms of the new SDW quasiparticle operators, the Hubbard Hamiltonian takes
the form
\begin{equation}
H={\sum_{{\bf k},\sigma}}^{\prime}~{\Bigg\{E^+_{\bf k}\Big(a^\dagger_{{\bf k},
\sigma}a_{{\bf k},\sigma}-b^\dagger_{{\bf k},\sigma}b_{{\bf k},\sigma}\Big)+
E^-_{\bf k}\Big(e^\dagger_{{\bf k},\sigma}e_{{\bf k},\sigma}-f^\dagger_{{\bf k}
,\sigma}f_{{\bf k},\sigma}\Big)\Bigg\}}
\label{Hubbard0}
\end{equation}
where
\begin{equation}
E^{\pm}_{\bf k}=\sqrt{(\epsilon^{\pm}_{\bf k})^2+\Delta^2},\qquad
\epsilon^{\pm}_{\bf k}=-4t\nu_{\bf k}\pm t', \qquad\nu_{\bf k}={\cos{k_x}+
\cos{k_y}\over 2}
\label{Endisp}
\end{equation}
and we set the lattice constant $a_0 =1$.
Here and below, the prime in the summation sign indicates that the
summation is restricted to the magnetic Brillouin zone, i.e. to momenta where
$\nu_{\bf k}>0$. For a two--layer system we obtain two
pairs of conduction (described by the $a$ and $e$) and valence ($b$ and $f$)
operators. The energy dispersions for each pair are shifted by ${\bf Q}=(\pi,
\pi)$. The direct gap $2 \Delta$ between the bands is determined by a
self--consistency equation for the staggered magnetization and reduces to
$2\Delta=U$ in the strong coupling limit.

The interaction between the fermionic current and the vector potential of
light, which is relevant for two--magnon scattering at transferred photon
frequencies small compared to the Mott--Hubbard gap, has the form
\begin{equation}
H_j = -\frac{e}{\hbar c} \sum_{\bf q}{\bf j}_{\bf q}\cdot{\bf A}_{-{\bf q}}\,\,
 .
\label{light}
\end{equation}
${\bf j}_{\bf q}$ is the current operator whose momentum can be safely set to
zero since the velocity of light is several orders of magnitude larger than the
Fermi velocity. To lowest order in $t/U$ the components of the current operator
are given by
\begin{equation}
j_{{\bf q}=0}^{\alpha}={\sum_{{\bf k},\sigma}}^{'}{\partial\epsilon_{\bf k}
\over\partial k_{\alpha}}\Bigg[a^\dagger_{{\bf k},\sigma}b_{{\bf k},\sigma}+
e^\dagger_{{\bf k},\sigma}f_{{\bf k},\sigma}+h.c.\Bigg]\,\, .
\label{jq0}
\end{equation}
 We see that to lowest order in $t/U$ the interaction with
light only leads to excitations of quasiparticles between the valence and
conduction bands of each pair. Excitations within each band are of higher order
in $t/U$ and are therefore neglected in our present strong coupling theory. It
is noteworthy that
the fermionic current rewritten in terms of quasiparticles which decouple
the Hamiltonian into two separate terms (i.e., $\alpha$ and $\beta$ operators
introduced in (\ref{rotation})),
 contains only densities of these fermionic quasiparticles. As a result,
there are no terms in $j_{{\bf q}=0}$ which would correspond to
excitations between the valence band of one pair
and the conduction band of the other pair
(e.g. $a^\dagger_{{\bf k},\sigma}f_{{\bf k},\sigma}$)~\cite{co}.

We now present the result for the Hamiltonian which describes the
magnon--fermion interaction. A systematic way to derive this Hamiltonian is to
extend the Hubbard model to a large number of $n=2S$ orbitals at a given site,
and use a $1/S$ expansion~\cite{Affleck,Chu-Mus}. The resulting spin--wave
spectrum is that for a spin $S$ antiferromagnet. In this Section, we will
consider only noninteracting spin waves, i.e. we will keep only the leading
linear term in the $1/S$ expansion. To simplify notations we will, however, not
keep the overall factors of $S$ in the formulae and thus present the results
for $n=2S=1$.

The spin--wave excitation spectrum of the two--layer antiferromagnet consists
of one doubly degenerate branch with the dispersion ~\cite{Matsuda}
\begin{equation}
\Omega_1({\bf q})=2 J_1 \sqrt{(1-\nu_{\bf q})(1+\nu_{\bf q}+J_2/2J_1)}
\label{mag-disp}
\end{equation}
for momenta ${\bf q}$ in the first Brillouin zone. The dynamic transverse spin
susceptibility has poles at $\Omega=\Omega_1({\bf q})$ and $\Omega=\Omega_1
({\bf q}+{\bf Q})$. For a single--layer antiferromagnet, $\Omega_1({\bf q})=
\Omega_1({\bf q}+{\bf Q})$, and the two poles are indistinguishable. For
two--layer systems, however, $\Omega_1({\bf q})$ and $\Omega_1({\bf q}+{\bf
Q})$ are different, and it is convenient to introduce two types of magnon
operators $m_{\bf q}$ and $n_{\bf q}$ with the dispersions $\Omega_1({\bf q})$
and $\Omega_2({\bf q})=\Omega_1({\bf q}+{\bf Q})$.

The electron--magnon interaction Hamiltonian can be obtained in the same way as
for a single--layer antiferromagnet ~\cite{Chu95}: it is uniquely defined by
the requirement that it should reproduce the forms of the dynamic spin
susceptibilities, both with zero momentum transfer and with momentum transfer
${\bf Q}$. The susceptibilities are presented in Appendix~\ref{der-susc}, and
the interaction Hamiltonian which reproduces them has the form
\begin{eqnarray}
&&H_{el-mag}={\sum_{\bf k}}^{\prime}\sum_{{\bf q},\sigma}\Bigg[a^\dagger_
{{\bf k+q},\sigma}a_{{\bf k},-\sigma} m^\dagger_{-{\bf q}}\Gamma_{-+}({\bf k},
{\bf q})+e^\dagger_{{\bf k+q},\sigma}e_{{\bf k},-\sigma}m^\dagger_{-{\bf q}}
\Gamma_{--}({\bf k},{\bf q}) \nonumber \\
& &+b^\dagger_{{\bf k+q},\sigma}b_{{\bf k},-\sigma}m^\dagger_{-{\bf q}}
\Gamma_{++}({\bf k+q})+f^\dagger_{{\bf k+q},\sigma}f_{{\bf k},-\sigma}
m^\dagger_{-{\bf q}}\Gamma_{+-}({\bf k,q})+a^\dagger_{{\bf k+q},\sigma}
e_{{\bf k},-\sigma}n^\dagger_{-{\bf q}}\Psi_{-+}({\bf k,q}) \nonumber \\
&&+e^\dagger_{{\bf k+q},\sigma}a_{{\bf k},-\sigma}n^\dagger_{-{\bf q}}
\Psi_{--}({\bf k,q})+f^\dagger_{{\bf k+q},\sigma}b_{{\bf k},-\sigma}
n^\dagger_{-{\bf q}}\Psi_{+-}({\bf k,q})+b^\dagger_{{\bf k+q},\sigma}
f_{{\bf k},-\sigma}n^\dagger_{-{\bf q}}\Psi_{++}({\bf k,q}) \nonumber \\
&& + (a^\dagger_{{\bf k+q},\sigma}b_{{\bf k},-\sigma}+e^\dagger_{{\bf k+q},
\sigma}f_{{\bf k},-\sigma})m^\dagger_{-{\bf q}}\Phi_-({\bf k,q})+b^\dagger_
{{\bf k+q},\sigma}a_{{\bf k},-\sigma}+f^\dagger_{{\bf k+q},\sigma}e_{{\bf k},-
\sigma})m^\dagger_{-{\bf q}}\Phi_+({\bf k,q}) \nonumber \\
& &+ (a^\dagger_{{\bf k+q},\sigma}f_{{\bf k},-\sigma}+e^\dagger_{{\bf k+q},
\sigma}b_{{\bf k},-\sigma})n^\dagger_{-{\bf q}}\Xi_-({\bf k,q})+(f^\dagger_
{{\bf k+q},\sigma}a_{{\bf k},-\sigma}+b^\dagger_{{\bf k+q},\sigma}e_{{\bf k},-
\sigma})n^\dagger_{-{\bf q}}\Xi_+({\bf k,q})   \nonumber \\
& &+h.c. \Bigg] \,\, .
\label{intH}
\end{eqnarray}
The vertex functions in Eq. \ref{intH} are
\begin{eqnarray}
\Phi_{\pm}({\bf k,q})&=&{\sqrt{2}~\Delta}\Big[\eta_{\bf q}\pm\bar{\eta}_
{\bf q}\Big],\qquad \Xi_{\pm}({\bf k,q})={\sqrt{2} \Delta}\Big[
\bar{\xi}_{\bf q} \pm \xi_{\bf q} \Big], \nonumber \\
\Gamma_{-+,--}({\bf k,q})&=&{1\over \sqrt{2}}
\Big[-\eta_{\bf q}(\epsilon^{\pm}_
{{\bf k+q}}+\epsilon^{\pm}_{\bf k})+
\bar{\eta}_{\bf q}(\epsilon^{\pm}_{\bf k+q}-
\epsilon^{\pm}_{\bf k})\Big], \nonumber \\
\Gamma_{++,+-}({\bf k,q})&=&{1\over \sqrt{2}}
\Big[\eta_{\bf q}(\epsilon^{\pm}_
{{\bf k+q}}+\epsilon^{\pm}_{\bf k})+
\bar{\eta}_{\bf q}(\epsilon^{\pm}_{\bf k+q}-
\epsilon^{\pm}_{\bf k})\Big], \nonumber \\
\Psi_{-+,--}({\bf k,q})&=&{1\over 2\sqrt{S}}\Big[- \xi_{\bf q}(\epsilon^{\pm}_
{\bf k+q}+\epsilon^{\mp}_{\bf k})+\bar{\xi}_{\bf q}(\epsilon^{\pm}_{\bf k+q}-
\epsilon^{\mp}_{\bf k})\Big],\,\, \nonumber \\
\Psi_{++,+-}({\bf k,q}) &=& {1\over 2\sqrt{S}}\Big[\xi_{\bf q}(\epsilon^{\pm}_
{\bf k+q}+\epsilon^{\mp}_{\bf k})+\bar{\xi}_{\bf q}(\epsilon^{\pm}_{\bf k+q}-
\epsilon^{\mp}_{\bf k})\Big]\,\, ,
\label{vertexfunctions}
\end{eqnarray}
where, e.g. the upper signs are for $\Gamma_{-+}$ and the lower ones for
$\Gamma_{--}$, and
\begin{eqnarray}
\eta_{\bf q}&=&{1\over\sqrt{2}}\left({1-\nu_{\bf q}\over 1+\nu_{\bf q}+
J_2/2J_1}\right)^{1/4}, \quad  \bar{\eta}_{\bf q}={1\over\sqrt{2}}\left({1+
\nu_{\bf q}+J_2/2J_1\over 1-\nu_{\bf q}}\right)^{1/4}, \nonumber \\
\xi_{\bf q}&=&{1\over\sqrt{2}}\left({1+\nu_{\bf q}
\over 1-\nu_{\bf q}+J_2/2J_1} \right)^{1/4}, \quad {\bar\xi}_{\bf q}
={1\over\sqrt{2}}\left({1-\nu_{\bf q}+J_2/2J_1\over 1+\nu_{\bf q}}
\right)^{1/4} \quad \,\, .
\label{etaandxi}
\end{eqnarray}
As for the case of a single--layer, the vertices which involve fermions from
both, conduction and valence bands are of order $U$ whereas the vertices
involving only valence or only conduction band fermions are of order $t$ and
in the non--resonant regime can be omitted in the calculations to lowest order
in $t/U$.

We now have all necessary tools to calculate the Raman matrix element $M_R$ in
a diagrammatic technique. A simple experimentation shows that, just as in the
case of a single layer, there are three diagrams which contribute to $M_R$ to
leading order in $t/U$ (see Fig. \ref{non-res-diag}). One has to keep in mind,
though, that the calculation of these diagrams now involves four bands and
two different type of magnons. The integration over the internal frequencies in
the diagrams of Fig. \ref{non-res-diag} is straightforward, and we obtain:
\begin{enumerate}
\item for the diagram in Fig. \ref{non-res-diag}a

\begin{equation}
M^{(1,2)}_{R_1}=\pm 4~\Big(\lambda_{1,2}^2({\bf q})+\mu_{1,2}^2({\bf q})\Big)\,
{1\over N}{\sum_{{\bf k},\alpha,\beta}}^{\prime}\Bigg({\partial\epsilon_{\bf k}
\over\partial k_\alpha}e_{i\alpha}\Bigg)\Bigg({\partial\epsilon_{{\bf k-q}}
\over\partial k_\beta}e_{f\beta}\Bigg) \ {4\Delta\over 4\Delta^2-\Omega^2}
\label{MR1}
\end{equation}
where the upper sign is for $M^{(1)}_{R_1}$, and the lower sign for
$M^{(2)}_{R_1}$, the indices $1$ and $2$ indicate whether the final state
contains two magnons of type $m_{\bf q}$ or of type $n_{\bf q}$, respectively,
and $\Omega$ is a frequency equal to $\omega_i$ or $\omega_f$ which are
indistinguishable to leading order in $t/U$;

\item for the diagram in Fig. \ref{non-res-diag}b

\begin{equation}
M^{(1,2)}_{R_2}=-4~\lambda_{1,2}({\bf q})\,\mu_{1,2}({\bf q})\,{1\over N}
{\sum_{{\bf k},\alpha,\beta}}^{\prime}\Bigg({\partial\epsilon_{\bf k}\over
\partial k_\alpha}e_{i\alpha}\Bigg)\Bigg({\partial\epsilon_{\bf k}\over\partial
k_\beta}e_{f\beta}\Bigg) \ \Bigg[{8\Delta\over 4\Delta^2-\Omega^2}+{8\Delta
(4\Delta^2+\Omega^2)\over (4\Delta^2-\Omega^2)^2}\Bigg] \,\, ;
\label{MR2}
\end{equation}
\item for the diagram in Fig. \ref{non-res-diag}c

\begin{equation}
M^{(1,2)}_{R_3}=4~\lambda_{1,2}({\bf q})\;\mu_{1,2}({\bf q})\,{1\over N}
{\sum_{{\bf k},\alpha,\beta}}^{\prime}\Bigg({\partial\epsilon_{\bf k}\over
\partial k_\alpha}e_{i\alpha}\Bigg)\Bigg({\partial\epsilon_{\bf k}\over\partial
k_\beta}e_{f\beta}\Bigg) \ {8\Delta(4\Delta^2+\Omega^2)\over (4\Delta^2-
\Omega^2)^2}\,\, .
\label{MR3}
\end{equation}
\end{enumerate}

As before, ${\bf e}_{i,f}$ are the polarization vectors of the incident and
scattered photons, respectively, and the coefficients $\mu_{1,2}({\bf q})$ and
$\lambda_{1,2}({\bf q})$ are defined as
\begin{eqnarray}
&\sqrt{2}\mu_1({\bf q})=\bar{\eta}_{\bf q}+\eta_{\bf q} , & \qquad \sqrt{2}
\lambda_1({\bf q})=\bar{\eta}_{\bf q}-\eta_{\bf q}, \nonumber \\
&\sqrt{2}\mu_2({\bf q})=\bar{\xi}_{\bf q}+\xi_{\bf q} , & \qquad \sqrt{2}
\lambda_2({\bf q})=\bar{\xi}_{\bf q}-\xi_{\bf q} \,\, .
\end{eqnarray}
Simple algebra yields the relations
\begin{equation}
\mu_{1,2}({\bf q})=\Bigg[{1\over 2}\Bigg({4J_1+J_2\over2\Omega_{1,2}({\bf
q})}+1
\Bigg)\Bigg]^{1/2}, \quad \lambda_{1,2}({\bf q})=
{4J_1 \nu_{\bf q} \pm J_2\over \mid 4J_1 \nu_{\bf q}\pm J_2 \mid} \
\Bigg[{1\over 2}\Bigg({4J_1+J_2\over 2\Omega_{1,2}({\bf q})}-1
\Bigg)\Bigg]^{1/2}\,\, .
\label{Bog-coef}
\label{lm}
\end{equation}
Comparing these results with those in Ref.~\cite{Chu95} we observe that the
form of Eqs. (\ref{MR1})--(\ref{MR3}) is the same for a single-- and a
double--layer system. The information about the coupling between the layers is
therefore only contained in the coherence factors $\mu_{1,2}({\bf q})$ and
$\lambda_{1,2}({\bf q})$. This is a direct consequence of the fact that the
photons' vector potential only couples to the in--plane fermionic current.

Finally, we use the relations
\begin{eqnarray}
{1\over N}{\sum_{{\bf k},\alpha,\beta}}^{\prime}\Big({\partial\epsilon_{\bf k}
\over\partial k_\alpha}e_{i\alpha}\Big)\Big({\partial\epsilon_{\bf k-q}\over
\partial k_\beta}e_{f\beta}\Big)&=& t^2\Big[e_{ix}e_{fx}\cos{q_x}+e_{iy}e_{fy}
\cos{q_y}\Big], \nonumber \\
{1\over N}{\sum_{{\bf k},\alpha,\beta}}^{\prime}\Big({\partial\epsilon_{\bf k}
\over\partial k_\alpha}e_{i\alpha}\Big)\Big({\partial\epsilon_{\bf k}\over
\partial k_\beta}e_{f\beta}\Big)&=&t^2\Big[e_{ix}e_{fx}+e_{iy}e_{fy}\Big]\,\, ,
\end{eqnarray}
substitute them into Eqs. (\ref{MR1})--(\ref{MR3}), and obtain for the total
Raman matrix element $M_{R}^{(1,2)}=M_{R_1}^{(1,2)}+M_{R_2}^{(1,2)}+
M_{R_3}^{(1,2)}$
\begin{eqnarray}
M^{(1,2)}_R& =-8t^2\displaystyle{2\Delta\over 4\Delta^2-\Omega^2} &
\Big[2\lambda_{1,2}({\bf q})\mu_{1,2}({\bf q})(e_{ix}e_{fx}+e_{iy}e_{fy}) \mp
\nonumber \\
& & \quad \Big(\lambda_{1,2}^2({\bf q})+\mu_{1,2}^2({\bf q})\Big)(e_{ix}e_{fx}
\cos{q_x}+e_{iy}e_{fy}\cos{q_y})\Big] \,\, .
\label{MR}
\end{eqnarray}

Notice that $M^{(1)}_R({\bf q})= M^{(2)}_R({\bf q+Q})$. Since for the
calculation of the Raman intensity we have to integrate over the whole
Brillouin zone we can restrict our consideration to only one type of magnons,
and just multiply $M_R$ by $\sqrt{2}$.

We now change tracks and compute the matrix element $M_{R}$ within the
Loudon--Fleury theory \cite{Fle68}, i.e. we assume that the spins interact via
the Heisenberg Hamiltonian
\begin{equation}
H=J_{1}\sum_{<i,j>,\alpha}{\bf S}_{\alpha,i}\cdot{\bf S}_{\alpha,j}+J_{2}
\sum_{i}{\bf S}_{1,i}\cdot{\bf S}_{2,i}\,\, .
\label {H1}
\label{hamilt}
\end{equation}
As before, $\alpha=1,2$, and the scattering of light is described by the
Loudon--Fleury Hamiltonian
\begin{equation}
H_{L-F}=\Lambda\sum_{j,\delta,\alpha} ~\left(e_{ix}e_{fx}{\bf S}_{j,\alpha}
\cdot{\bf S}_{j+\delta_x,\alpha}+e_{iy}e_{fy}{\bf S}_{j,\alpha}\cdot{\bf S}_{j+
\delta_y,\alpha}\right)\,\, .
\label{RH}
\end{equation}
Here, $\Lambda$ is a coupling constant and $\delta=(\delta_x,\delta_y)$ is a
vector to nearest neighbors sites in a plane. Observe that this scattering
Hamiltonian has the same form as for a single layer. This is again a
consequence of the fact that the light only couples to the in--plane fermionic
current. Following the standard procedure, the spin operators are now
transformed to boson operators via the conventional Holstein--Primakoff or
Dyson--Maleev expansions. The easiest way to proceed is to introduce just one
Bose field with momentum in the full first Brillouin zone, and then perform a
unitary rotation to magnon operators which diagonalize the quadratic form for a
two--layer Heisenberg antiferromagnet. As for a single plane, the
transformation to magnon operators involves the same coefficients
$\lambda_{1}({\bf q})$ and $\mu_{1}({\bf q})$ as in Eq. (\ref{lm}).

Retaining only the term in the scattering Hamiltonian which contains two magnon
creation operators, we obtain for the Loudon--Fleury matrix element
\begin{eqnarray}
M_{R_{LF}}&=&-\Lambda \Big[2\lambda_1({\bf q})\mu_1({\bf q})(e_{ix}e_{fx}+
e_{iy}e_{fy})- \nonumber \\
& & \quad \Big(\lambda_{1}^2({\bf q})+\mu_{1}^2({\bf q})\Big)(e_{ix}e_{fx}
\cos{q_x}+e_{iy}e_{fy}\cos{q_y})\Big]\,\, .
\label{MRLF}
\end{eqnarray}
Comparing the two expressions for $M_R$, Eq. \ref{MR} and Eq. \ref{MRLF}, we
see that they are identical provided we identify the coupling constant
\begin{equation}
\Lambda=8\sqrt{2}t^2\left[{2\Delta\over 4\Delta^2-\Omega^2}\right]\,\, .
\end{equation}
This concludes our derivation of the Loudon--Fleury Hamiltonian for a
two--layer system.

Before we proceed with the calculations of the Raman intensity, we would like
to comment on the form of the Loudon--Fleury Hamiltonian. In some
phenomenological theories for a single layer, the interaction Hamiltonian
between light and spin degrees of freedom is written as
\begin{equation}
H_{LF} = {\Lambda}\sum_{j,\delta} P({\bf e}_{i}, {\bf e}_{f},\delta)
{\bf S}_{\alpha,j}\cdot{\bf S}_{\alpha,j+\delta}
\end{equation}
where
\begin{equation}
P({\bf e}_{i},{\bf e}_{f},\delta)=\left[{1\over 2}{\bf e}_{i}\cdot{\bf e}_{f}-
(\delta\cdot{\bf e}_{i})(\delta\cdot{\bf e}_{f})\right]\,\, .
\end{equation}
This formula is obtained from (\ref{RH}) if the term which describes scattering
in $A_{1g}$ geometry is neglected. For a single layer, this procedure is
legitimate as  the scattering Hamiltonian in $A_{1g}$ geometry commutes with
the Heisenberg Hamiltonian, and consequently there is no $A_{1g}$ scattering.
However, for a two--layer system, more care is needed as the Heisenberg
Hamiltonian now contains an extra term with an interlayer coupling, which
does not
commute with $H_{LF}$. As a result, if we want to rewrite the Loudon--Fleury
Hamiltonian for two layers using the projection operator $P$, we necessarily
have to introduce an extra term which contains spins from two {\it different}
planes. Specifically, the Loudon--Fleury Hamiltonian expressed in terms of
projection operators should have the form
\begin{equation}
H_{LF}=\Lambda_{B1} ~\sum_{j,\delta,\alpha}P({\bf e}_i,{\bf e}_{f},\delta)
{\bf S}_{\alpha,j}\cdot{\bf S}_{\alpha,j+\delta}+\Lambda_{A1} ~\sum_{j}P({\bf
e}_i,{\bf e}_f,0){\bf S}_{1,j}\cdot{\bf S}_{2,j}
\end{equation}
where $P({\bf e}_i,{\bf e}_f,\delta)$ is the same as before. Here, the first
term describes scattering in $B_{1g}$ geometry, while the second term (which
couples spins from different layers) contributes to $A_{1g}$ scattering.
Comparing the two forms of the Loudon--Fleury Hamiltonians, we obtain for the
two coupling constants
\begin{equation}
\Lambda_{A1}=4t^2~\left[{2\Delta\over 4\Delta^2-\Omega^2}\right]{J_2\over J_1},
\qquad \Lambda_{B1}=-8t^2~\left[{2\Delta\over 4\Delta^2-\Omega^2}\right] \,\, .
\end{equation}
The two forms of the Loudon--Fleury Hamiltonian indeed yield the same Raman
matrix element as in Eq.~(\ref{MRLF}).

\section{The Raman Intensity in the Non--Resonant Regime}
\label{raman-int}
\subsection{The Non-Interacting Case}

The Raman scattering cross section is proportional to the Golden Rule
transition rate \cite{Hay78}
\begin{equation}
R={8\pi^3 e^4\over\hbar^3 V^2\omega_i\omega_f}~\sum {\mid M_R\mid^2\delta
(\hbar\omega_i-\hbar\omega_f+\epsilon_i-\epsilon_f )}
\label{raman}
\end{equation}
where $i$ and $f$ are the initial and final states of the system,
$\epsilon_{i,f}$ are the corresponding energies, and the
summation runs over all possible initial and final electronic states.
 Let us first neglect final
state magnon--magnon interaction. Then, $\epsilon_{i}-\epsilon_{f}=2
\Omega_1({\bf q})$, and using (\ref{MR}) we obtain for the Raman intensity in
the $A_{1g}$ and $B_{1g}$ channels (dropping an identical overall prefactor)
\begin{eqnarray}
I_{A_{1g}}(\Omega)&\propto& 2\sum_{\bf q}\left(M_R^{(1)}\right)^2\> \delta
\Big(\hbar\omega_i-\hbar\omega_f-2\Omega_1({\bf q})\Big) \nonumber \\
&=&2\Lambda^2~\sum_{\bf q}\left(\frac{J_2}{4 J_1}\right)^2~ \left({2J_1(1-
\nu_{\bf q})\over\Omega_1({\bf q})}\right)^2 \> \delta\Big(\hbar\omega_i-
\hbar\omega_f-2\Omega_1({\bf q})\Big)\, , \nonumber \\
I_{B_{1g}}(\Omega)&\propto&2\sum_{\bf q}\left(M_R^{(1)}\right)^2\> \delta
\Big(\hbar\omega_i-\hbar\omega_f-2\Omega_1({\bf q})\Big) \nonumber \\
&=&2\Lambda^2~\Big(1+{{J}_2\over 4{J}_1}\Big)^2\sum_{\bf q}\left({2J_1
\tilde{\nu}_{\bf q}\over\Omega_1({\bf q})}\right)^2 \> \delta\Big(\hbar
\omega_i-\hbar\omega_f-2\Omega_1({\bf q})\Big)
\label{A1g}
\end{eqnarray}
where $\Omega=\omega_i-\omega_f$ and we have defined $\tilde{\nu}_{\bf q}=
(\cos{q_x}-\cos{q_y})/2$. We see that the Raman intensity in the $A_{1g}$
channel is proportional to $(J_2/J_1)^2$ and thus vanishes with vanishing
$J_2$. In $B_{1g}$ geometry, the changes in $I_{B_{1g}}$ imposed
by the interlayer coupling are minor: the form of the matrix element is
preserved
and the only changes appear in the prefactor and in the magnon energy
dispersion $\Omega_1({\bf q})$. The momentum sums in Eq.~(\ref{A1g}) can be
conveniently reduced to complete elliptic integrals. The resulting expressions
for the intensities are, however, rather involved; they are collected in
Appendix \ref{raman-int-app}. Here we discuss only the main features of the
Raman spectrum.

Our key observation is the following: The magnon energy $\Omega_1({\bf q})$ is
gapless at the zone center ${\bf q}=0$, and has a gap, $\Omega_1({\bf Q})=2
\sqrt{{J}_2{J}_1}$ at ${\bf q}={\bf Q}$. Then, for $\Omega=\omega_i-\omega_f<2
\Omega_1({\bf Q})$, only magnons with momentum near ${\bf q}\approx 0$ can be
excited. The numerators in both scattering geometries vanish at ${\bf q}=0$,
and it is not difficult to show that the contributions from the ${\bf q}
\approx 0$ region yield $I(\Omega)\propto\Omega^3$ (see also below). For
$\Omega>2\Omega_1({\bf Q})$, however, also magnons with ${\bf q}\approx
{\bf Q}$ can be excited. In $B_{1g}$ geometry, the numerator in $I_{B_{1g}}$
contains the factor $\tilde{\nu}_{\bf q}^2$ which vanishes at ${\bf q}={\bf Q}$
such that the opening of a new scattering channel does not cause substantial
changes in the intensity which still scales as $\Omega^3$. However, in $A_{1g}$
geometry the numerator in $I_{A_{1g}}$ at ${\bf q}={\bf Q}$ is just a positive
constant. In this situation, the scattering intensity changes drastically at
$\Omega=2\Omega_1({\bf Q})$: using Eq.~(\ref{A1g}), we find
\begin{equation}
\Delta I_{rel}={I_{A_{1g}}(2\Omega_1({\bf Q})+\delta\omega)-I_{A_{1g}}(2
\Omega_1({\bf Q})-\delta\omega)\over I_{A_{1g}}(2\Omega_1({\bf Q})-\delta\omega
)}={256 J_1^4\over(\Omega_1({\bf Q}))^4}\,\, .
\end{equation}
For ${J}_2/{J}_1=0.1$ we obtain $\Delta I_{rel}\sim 1600$, i.e. the
enhancement of the $A_{1g}$ signal at the threshold frequency is very strong.

Moreover, the value of the intensity in the $A_{1g}$ channel right above the
threshold, $I_{A_{1g}}\propto J_2^2/\Omega_1({\bf Q})\propto J_1
(J_2/J_1)^{3/2}$, has the same order of magnitude as the intensity in the
$B_{1g}$ channel, $I_{B_{1g}}\propto(\Omega_1({\bf Q}))^3\propto
(J_2/J_1)^{3/2}$. We found that for all reasonable values for ${J}_2/{J}_1$ the
ratio of intensities is $\sim 1.7-1.9$. In other words, if the Loudon--Fleury
approximation is applicable, and if one claims to observe the two--magnon
profile in $B_{1g}$ geometry at around $\Omega_1({\bf Q})$, one should also
observe, in a two--layer system, the signal of an even larger intensity in the
$A_{1g}$ geometry.

The intensities for the $A_{1g}$ and $B_{1g}$ geometries without final state
interaction are plotted in Fig.~\ref{B1g-intens} for two different values of
${J}_2/{J}_1$. There are unphysical singularities in both intensities at the
maximum magnon energy, but just as in the case of a single layer, they are
artifacts of neglecting interactions between magnons. We will see in Sec.
\ref{interacting-case} that once an interaction is included, the unphysical
singularities are removed.

\subsection{The Interacting Case}
\label{interacting-case}
We now analyze how the two--magnon profile changes when the interaction between
magnons is included. Fist of all, the magnon--magnon interaction renormalizes
the spin--wave spectrum. To leading order in $1/S$, which we only consider
here, this renormalization can be absorbed into the renormalization of the
exchange couplings
\begin{eqnarray}
J_1&\rightarrow &J_1\left(1+{r\over 2S}\right)\, ,\qquad r=1-{1\over N}
\sum_{\bf q}{(1-\nu_{\bf q})\left(4J_1(1+\nu_{\bf q})+J_2\right)\over
{2\Omega_1({\bf q})}}\, , \nonumber \\
J_2&\rightarrow &J_2\left(1+{r'\over 2S}\right)\, ,\qquad r'=1-{1\over N}
\sum_{\bf q}{4J_1(1-\nu_{\bf q})\over{2\Omega_1({\bf q})}}\,\, .
\label{renorm}
\end{eqnarray}
where the momentum sums run over the whole first Brillouin zone (notice that we
defined $\Omega_1$ without a factor of $2S$). This renormalization comes from
one--loop diagrams (Oguchi corrections~\cite{Ogu60}). Beyond the leading order
in $1/S$, one has to solve Eq. (\ref{renorm}) self--consistently and also
include corrections with higher number of loops. Numerically, however, it turns
out that the dominant correction, at least to order $1/S^2$, still comes from
one--loop diagrams~\cite{Chu95a}. In other words, the actual magnon dispersion
nearly preserves the same form as in linear spin wave theory, but contains
renormalized coupling constants $J_{1,2}$. Below we will assume that this
renormalization is already included into the definitions of $J_{1,2}$ and
neglect it in our further consideration.

Strictly speaking, to justify this approximation for the calculation of the
effects due to a final state magnon--magnon interaction, we also have to prove
that the dominant renormalization of the four--magnon interaction vertex can be
absorbed into the same renormalization of the exchange integrals. We did not
perform explicit $1/S$ calculations for the vertex. However, as the vertex
itself has a factor $1/S$ in comparison to the magnon frequency, and all
calculations involving magnon--magnon scattering will be performed only to the
leading order in $1/S$, then whether to use a bare or a renormalized $J_{1,2}$
in the vertex is beyond the accuracy of our calculations. For simplicity, we
will henceforth use renormalized values of $J_{1,2}$ everywhere.

We now consider in detail the renormalization of the two--magnon profile due to
multiple scattering of two magnons. The magnon--magnon vertices can be
immediately obtained from the Heisenberg Hamiltonian by applying e.g. the
Holstein--Primakoff transformation to boson operators and a subsequent
canonical transformation to magnon operators which diagonalize the quadratic
part of the spin wave Hamiltonian. A detailed study of the effects due to
magnon--magnon interaction in a single--layer antiferromagnet was already
performed by Canali and Girvin \cite{Can92}, and we follow here their line of
reasoning. To leading order in $1/S$, we can restrict ourselves to the
scattering process which conserves the number of magnons. The effective
scattering Hamiltonian then takes the form
\begin{equation}
H_{mag-mag}={1\over N}\sum_{{\bf k,q}}~V({\bf k,q})\alpha^\dagger_{\bf q}
\beta^\dagger_{-{\bf q}}\alpha_{\bf k}\beta_{-{\bf k}}
\label{interact-2}
\end{equation}
with
\begin{equation}
V({\bf k,q})= A{B_{\bf k}\ B_{\bf q}\over\Omega_1({\bf k})\Omega_1(
{\bf q})}- ~B_{\bf q-k}\Bigg[{A^2\over\Omega_1({\bf k})\Omega_1
({\bf q})}+1\Bigg] \,\, ,
\label{interaction}
\end{equation}
and
\begin{equation}
A=(J_2+4J_1)/2\, ,\quad B_{\bf k}=(J_2+4J_1\nu_{\bf k})/2\,\, .
\end{equation}
In terms of $A$ and $B_{\bf k}$ the magnon dispersion is given by $\Omega_1
({\bf k})=\sqrt{A^2-B_{\bf k}^2}$.

In order to find the full vertex function for repeated two--magnon scattering
we need to sum an infinite series of ladder diagrams. In $B_{1g}$ geometry, the
``side'' vertices from the electron--photon coupling scale as ${\tilde\gamma}_{
\bf q}=(\cos q_x-\cos q_y)/2$, and it is easy to see that the only term in
Eq.~(\ref{interaction}) which contributes to scattering is the one with
$\nu_{\bf q-k}$. The evaluation of the ladder diagram series then proceeds
exactly in the same way as for a single--layer system \cite{Can92,Dav71}. The
analytical solution is presented in Appendix \ref{magn-int}.

The plots for the Raman intensity $I_{B_{1g}}$ for two values of $J_2/J_1$ are
shown in Fig.~\ref{B1g-int-fig1}. The unphysical singularity that we found in
the non--interacting case disappears, as expected, and we observe a pronounced
two--magnon peak. We see that with increasing $J_2/J_1$ the two--magnon peak
not only shifts to higher frequencies but that the amplitude of the signal also
slightly increases. The latter, however, is mainly due to the overall factor
$(1+J_2/4J_1)$ in the matrix element and a renormalization of the
Loudon--Fleury constant $\Lambda$ from magnon--magnon interactions. The shift
of the peak position towards higher frequencies can be understood in the simple
picture that the incoming photon flips two neighboring spins on the same layer.
This creates misaligned spin pairs and thereby increases the total energy of
the system. Evaluating the corresponding energy increase for a N\'eel state, we
obtain a two--magnon peak at $\Omega=3J_1(1+J_2/3J_1)$ which is roughly
consistent with what we find.

In $A_{1g}$ geometry, the solution of the ladder series is more difficult since
the ``side'' vertex  behaves as $\sim(1-\nu_{\bf q})/\Omega_1({\bf q})$ where
${\bf q}$ is the magnon momentum. At small ${\bf q}$, this vertex scales
linearly with ${\bf q}$ as a consequence of the Adler principle~\cite{Adler}:
the Raman matrix element includes the interaction between fermions and
Goldstone bosons, and this interaction should vanish at the points where the
magnon energy turns to zero. Because of the extra power of momentum in $M_R$,
the Raman intensity without final state interaction scales as $I_{A_{1g}}
\propto\Omega^3$ at very low frequencies, as mentioned before. However, the
form of the ``side'' vertex in $A_{1g}$ geometry, is {\it not} reproduced at
the magnon--magnon vertex, and we in fact have to solve a set of coupled
integral equations in order to get the result for the full $I_{A_{1g}}$. The
explicit expression for $I_{A_{1g}}$ is rather cumbersome, so we
present it in Appendix \ref{magn-int} and here discuss only the key features of
the solution.

As the ``side'' vertices for $A_{1g}$ are invariant under transformations of
the symmetry group $D_{4h}$ of the square lattice, we can restrict ourselves to
only that part of the scattering potential $V({\bf k,q})$, which has the same
symmetry, i.e.
\begin{equation}
V({\bf k,q})=-2J_1\Bigg[\nu_{\bf k}\nu_{\bf q}+{J_2\over 4J_1}+{J_2\over 4J_1}
\Big(1+{J_2\over 4J_1}\Big){(2J_1)^2(1-\nu_{\bf k})(1-\nu_{\bf q})\over
\Omega_{\bf q} \ \Omega_{\bf k}}\Bigg]\,\, .
\label{vsymm}
\end{equation}
We see that $V({\bf k,q})$ actually tends to a finite value for
${\bf k}={\bf q}=0$. The cubic frequency dependence of $I_{A_{1g}}(\Omega)$ is
therefore actually an artifact of neglecting the final state interaction.
When this
interaction is included, $I_{A_{1g}}$ scales {\it linearly} with $\Omega$ at
the lowest frequencies. We also found that the real part of the polarizability
has a logarithmic singularity at $\Omega=2\Omega_1({\bf Q})$. This singularity
gives rise to two effects: first, it makes $I_{A_{1g}}$ a continuous function
of frequency, in other words, eliminates a jump in the intensity at
$2\Omega_1({\bf Q})$. Second, it gives rise to a strong peak in $I_{A_{1g}}$ at
frequencies somewhat smaller than $2\Omega_1({\bf Q})$. Specifically, we found
that near $2\Omega_1({\bf Q})$, the dominant contribution to the intensity
comes from the third term in Eq.~(\ref{vsymm}), and $I_{A_{1g}}$ has the form
\begin{equation}
I_{A_{1g}}\propto J_2^2~\frac{\Omega^3_1({\bf Q})}{(1+R_1)^2+R^2_2}+
I^{\prime}_{A_{1G}}
\end{equation}
where
\begin{equation}
R_1=\frac{J_2}{2\pi\Omega_1({\bf Q})}~\ln{\frac{2\Omega_1({\bf Q})-\Omega}{2
\Omega_1({\bf Q})}}\hskip1.cm , \hskip1.cm R_2={1\over 4}{J_2\over J_1}\left({
\Omega_1({\bf Q})\over 2J_1}\right)^3\,\, .
\label{R_}
\end{equation}
$I^{\prime}_{A_{1g}}$ remains finite at $\Omega=2\Omega_1({\bf Q})$ and is
proportional to $ J^2_2/\Omega_1({\bf Q})$. We see that there exists a very
narrow peak in $I_{A_{1g}}$ located at $\Omega=2\Omega_1({\bf Q})(1-\exp
(-2\pi\Omega_1({\bf Q})/J_2))$. The intensity right at the peak is very high,
$I_{A_{1g}}\propto 1/\Omega^3_1({\bf Q})$. At small $J_2$, the peak position is
exponentially close to $2\Omega_1({\bf Q})$. However, at larger $J_2$, we found
numerically that the peak is actually located at frequencies significantly
smaller than $2\Omega_1({\bf Q})$. This last result agrees with the
calculations of the two--magnon absorption profile in Ref. \cite{Gruninger95}.

The solutions for $I_{A_{1g}}$ for two different values of $J_2/J_1$ are
graphically presented in Fig.~\ref{A1g-int-fig}. The $A_{1g}$ Raman spectra
have been evaluated on a finite lattice with $1000\times 1000$ lattice points.
A finite imaginary part $i\delta$ has been added to the energy denominators of
the spin wave propagators in Eqs. (\ref{P}), (\ref{R}), and (\ref{Q}) in
Appendix \ref{magn-int}. This allows to study the influence of damping on the
$A_{1g}$ two--magnon spectra. Without damping, the imaginary part of the
polarizability has a jump at $\Omega=2\Omega_1({\bf Q})$, and by the
Kramers--Kronig relation, the real part of polarizabiliy ($R_1$ term in
Eq. (\ref{R_})) necessarily has a logarithmic
singularity, see Fig.~\ref{A1g-int-fig}c.
 With damping, the singular behavior near the threshold frequency
is removed and the peak position is shifted closer to $2\Omega_1({\bf Q})$.
Note also that, as in $B_{1g}$ geometry, the divergence at twice the maximum
spin wave frequency is removed due to the final state magnon--magnon
interaction.

Finally, for the purpose of comparison with experiments, it is useful to
compute the ratio of the Raman intensities for $A_{1g}$ and $B_{1g}$ geometries
right at their peak positions. We found that this ratio is actually very small:
for ${J}_2/{J}_1=0.1$ it is about $0.009$ whereas for $J_2/J_1=0.3$, it is
$0.044$. In other words, though the $A_{1g}$ intensity at the peak is larger
than the intensity of the $B_{1g}$ signal at the same frequency, the overall
scale of the peak is only a few percent of the two--magnon peak in $B_{1g}$
geometry. We therefore have to conclude that in the non--resonant regime where
the Loudon--Fleury theory is applicable, the extra peak in $A_{1g}$ geometry
can hardly be separated from the background signal. We now consider what
happens in the resonant regime, i.e. when the incident photon frequency becomes
comparable to the Mott--Hubbard gap.

\section{The Raman Intensity in the Resonant Regime}
\label{resonant-case}

In our derivation of the Loudon--Fleury Hamiltonian, we have chosen the
diagrams to leading order in $t/U$ under the assumption that the energy of the
incoming and outgoing photons is much smaller than the energy gap between the
conduction and valence bands. Under these conditions, all denominators in the
diagrams Figs. \ref{non-res-diag}a--c were of order $U$ which in turn allowed
us to omit all diagrams with intraband scattering. The situation, however,
becomes different in the resonant regime where the energy of the incoming
photon comes close to the Mott--Hubbard gap $2\Delta \sim U$. Actually, most of
the experiments on two--magnon Raman scattering in parent high--$T_c$ compounds
have been done with visible light frequencies which are only $O(J)$ apart from
$2\Delta$.

It was shown in Ref. \cite{Chu95} that in the resonant regime, the diagrams
with intraband scattering are more relevant than those which contribute to the
Loudon--Fleury theory, and, moreover, the dominant contribution to Raman
scattering comes from just one diagram shown schematically in Fig.
\ref{res-diag}. This diagram yields a Raman matrix element
$M_R^{tr}=M^+_R+M^-_R$, where
\begin{equation}
M^{\pm}_R=8i~{1\over N}{\sum_{{\bf k},\alpha,\beta}}^{\prime}{\Big(
\displaystyle{\partial\epsilon_{\bf k}\over\partial k_\alpha}e_{i\alpha}\Big)
\Big(\displaystyle{\partial\epsilon_{\bf k-q}\over\partial k_\beta}e_{f\beta}
\Big)\Big[\mu_{\bf q}\epsilon^{\pm}_{\bf k-q}-\lambda_{\bf q}\epsilon^{\pm}_{
\bf k}\Big]^2\over\Big(\omega_i-2E^{\pm}_{\bf k}+i\delta\Big)\Big(\omega_i-
\Omega_1({\bf q})-E^{\pm}_{\bf k}-E^{\pm}_{\bf k-q}+i\delta\Big)\Big(\omega_f-2
E^{\pm}_{\bf k-q}+i\delta\Big)}\,\, .
\label{triple}
\end{equation}
One of the key consequences of considering the resonance regime is that there
exists a nonzero signal in $A_{1g}$ geometry even for a single--layer. Indeed,
the absence of the $A_{1g}$ signal in the Loudon--Fleury theory was related to
a particular form of the interaction Hamiltonian $H_{LF}$ which contained only
spin degrees of freedom. The inclusion of the intraband processes modifies the
form of the interaction Hamiltonian with light, in which case it no longer
commutes with the Heisenberg Hamiltonian even when ${\bf e}_i={\bf e}_f=({\hat
x}+{\hat y})/\sqrt{2}$.

The feature of the diagram in Fig. \ref{res-diag} which makes it dominant for
$B_{1g}$ scattering in the resonance regime is  that it allows all three
denominators to vanish simultaneously leading to a triple resonance enhancement
\cite{Chu95}. For two--layer systems, we should check  whether or not the rapid
variation of $I_{A_{1g}}$ near $2\Omega_1({\bf Q})$, can be enhanced when the
incident photon frequency is tuned right to the triple resonance value. We
performed computations analogous to those in \cite{Chu95} and found that there
is in fact no enhancement at the frequency threshold for $A_{1g}$ scattering
because the occurrence of the triple resonance requires that the fermionic
velocities at momenta ${\bf k}_0$ ($\omega_i=2E^{\pm}({\bf k}_0)$) and
${\bf k}_0+{\bf q}$, where ${\bf q}$ is the magnon momentum, be antiparallel to
each other. For ${\bf q}={\bf Q}$, we evidently have $\nabla_{\bf k}
E_{\bf k}^\pm\mid_{{\bf k}_0}=\nabla_{\bf k} E^\pm_{\bf k}\mid_{{\bf k}_0+{\bf
Q}}$, i.e. the two velocities are {\it parallel}. In this situation, the
integration over the fermionic momenta near ${\bf k}_0$ gives zero because all
poles lie in the same half--plane. We also performed more detailed calculations
by expanding the denominators up to second order around ${\bf k}_0$. This
actually makes the integral over ${\bf k}-{\bf k}_0$ finite, but still there is
no singularity in $M^{tr}_R$ at $\omega_i-2E_{{\bf k}_0}$, so we do not expect
any substantial enhancement of the Raman intensity in the $A_{1g}$ channel due
to a triple resonance.

Despite the absence of the enhancement, the diagram in Fig. \ref{res-diag} is
still relevant in the resonant regime simply because it contains three
denominators which all are $O(J)$. Since there is no resonant enhancement,
then, to first approximation, one can just set the denominator in
(\ref{triple}) to a constant and consider the basic structure of $M_R$ as
imposed by the interaction vertices between magnons and fermions. Performing
simple calculations, we obtained from (\ref{triple})
\begin{equation}
M^{tr}_R \propto \left (\nu_{\bf q}~\Omega_1({\bf q}) ~+~{J_1~J_2 \over 4} \,
 \frac{1 - \nu_{\bf q}}{\Omega_1({\bf q})} \right)  \,\, .
\label{trip}
\end{equation}
At small frequencies, the contributions to the Raman intensity come only
from magnon momenta near ${\bf q}=0$. We see from (\ref{trip}) that for these
momenta, $M^{tr}_R$ scales linearly with the magnon momentum, just as we found
in the Loudon--Fleury approximation. Clearly then, the full Raman intensity in
the absence of magnon--magnon interaction scales as $I_{A_{1g}}\propto\Omega^3$
at small frequencies. We studied the effects of the magnon--magnon interaction
and found that, as before, the bare form of the side vertex is not reproduced
in a perturbation theory for magnon--magnon scattering, and the finite state
interaction gives rise to a linear, rather than cubic frequency dependence of
$I_{A_{1g}}$. Moreover, as $M^{tr}_R$ does not contain $J_2$ as the overall
factor, it obviously gives a dominant contribution to $I_{A_{1g}}$. This in
turn implies that at $\Omega<2\Omega_1({\bf Q})$, the Raman intensity in a
double--layer system is roughly half of the intensity in a one--layer system.

For $\Omega>2\Omega_1({\bf Q})$, magnon momenta near ${\bf q}={\bf Q}$ also
contribute to $I(\Omega)$. It is not difficult to verify that this extra
contribution has the same dependence on $J_2/J_1$ as in the Loudon--Fleury
theory. Accordingly, if we consider only $M^{tr}_R$, we obtain qualitatively
the same form of the $A_{1g}$ intensity profile as in the Loudon--Fleury theory
-- the only difference is that now the intensities below and above the jump at
$2\Omega_1({\bf Q})$ are of the same order of magnitude. Our result for the
$A_{1g}$ intensity computed with $M^{tr}_R$ with final state interaction is
presented in Fig.~\ref{triple_fig}. Qualitatively, the intensity profile is the
same as in the Loudon--Fleury approximation, but the new features are a
substantial increase in the intensity above the threshold at $2\Omega_1({\bf
Q})$ and the flattening of the $A_{1g}$ intensity slightly above the threshold
frequency. There may also be a very narrow peak slightly below the threshold
frequency (just as we obtained in the Loudon--Fleury theory), which we do not
see because of a limited numerical accuracy. In any event, however, the
singular behavior at this peak is eliminated by damping.

The total matrix element for the $A_{1g}$ scattering is a sum of Loudon--Fleury
and triple resonance contributions. Without studying in detail the frequency
dependence of the denominator in (\ref{triple}) we cannot compare the overall
strength of $M^{LF}_R$ and $M^{tr}_R$. In general, in the absence of the
enhancement due to an actual triple resonance, the two contributions should
have the same order of magnitude ~\cite{Chu95}. Experimentally, however, the
overall intensity (and, to some extent, the form) of the $A_{1g}$ Raman profile
demonstrates a substantial dependence on the incident photon frequency.
Besides, as we noted above, the Loudon--Fleury result for the $A_{1g}$
intensity at the threshold is more than $1000$ times smaller than the $B_{1g}$
intensity at its maximum, while the experimental intensity ratio is about $40$
times smaller in the vicinity of the triple resonance in $B_{1g}$ geometry, and
even far smaller away from the resonance. It is therefore very likely that the
Loudon--Fleury contribution to the $A_{1g}$ intensity is just a minor
correction to the intensity given by the triple resonance diagram. Notice also
that $M^{LF}_R$ has exactly the same form as the second term in $M^{tr}_R$, and
its inclusion will just change the relative strength of the two terms in
(\ref{trip}).

\section{Discussion}
\label{discussion}
We first summarize our results. We considered in this paper two--magnon Raman
scattering in a two--layer Hubbard model at half--filling. We applied the SDW
formalism and derived diagrammatically the Loudon--Fleury Hamiltonian for the
interaction between light and spin degrees of freedom. We found that in a
two--layer system, the scattering in $A_{1g}$ geometry is finite already in the
Loudon--Fleury approximation. Without final state interaction, the intensity in
this channel scales as $\Omega^3$ at low frequencies. The magnon--magnon
interaction effects are numerically small, but nevertheless they change the
frequency dependence to $I_{A_{1g}}\propto\Omega$ at small frequencies.
Furthermore, there is a very strong resonance near $\Omega=4(J_1J_2)^{1/2}$,
when a second scattering channel opens up. At resonance, the amplitude of the
$A_{1g}$ signal is larger than the amplitude of the $B_{1g}$ signal at the same
frequency. We also argued that in the resonance regime relevant to experiments
on parent high-$T_c$ compounds,
there is no enhancement of the peak intensity in $A_{1g}$ geometry due to the
actual triple resonance. Nevertheless, the diagram with three resonant
denominators is dominant in this regime
 as it yields a finite $A_{1g}$ intensity even without $J_2$.

In Fig.~\ref{exp} we present the experimental data for the $A_{1g}$ Raman
intensity for the single--layer  $Sr_2CuO_2Cl_2$ and the double--layer
$YBa_2Cu_3O_{6.1}$ compounds~\cite{Blumberg95}. We see that at transferred
frequencies $\omega\geq 2000cm^{-1}$, the intensity profiles in the two
compounds are similar. The sharp peaks at the low energy tail of the
two--magnon band are due to resonant multi--phonon scattering that becomes
strongly enhanced for excitations close to $2\Delta$. Despite the overall
similarity of the two figures, there are clear differences at low frequencies.
The intensity in a single--layer compound continues to decrease at frequencies
smaller than the resonance frequencies for phonon scattering, while the
intensity for a two--layer compound flattens at frequencies somewhat larger
than the resonance frequencies for phonon scattering, and remains flat down to
the smallest measured frequencies.

At the moment, we do not understand the origin of the background contribution
to the scattering in $YBCO$, but it is unlikely that this background
contribution is related to scattering in a half-filled insulator. A more likely
possibility is that the background is due to the fact that the measured $YBCO$
compound has some finite amount of holes. In any event, however, we see that
the intensity flattens at about $1800 cm^{-1}$, and is rather flat at even
lower frequencies.

This behavior is consistent with our result for the resonant regime where the
experiments have been performed: the $A_{1g}$ intensity evaluated for $J_2=0.1
J_1$ flattens at about $1.8J_1 \sim 1800 cm^{-1}$, and is roughly two times
flatter at low frequencies than in a single--layer compound. Indeed, the theory
also predicts that there should be a jump in the intensity at the threshold
frequency. However, if we associate the onset of flattening with $1800cm^{-1}$,
we find that the jump occurs at about $1200 cm^{-1}$, i.e., right at the
frequencies where the Raman signal is presumably dominated by phonon
scattering, so there are little chances to observe this jump directly. We
therefore believe that $J_2\sim 0.1J_1$ is a reasonable though indirect
estimate of $J_2$.

Notice that $J_2\sim 0.1J_1\sim 150K$ is consistent with the estimate obtained
from the analysis of the NMR data in similar systems. More rigorously, we can
place the upper bound on possible values of $J_2$ because whatever the
interpretation of the low frequency measurements is, the data above
$2000cm^{-1}$ clearly show no influence of the interlayer coupling. This in
turn implies that in any event, the threshold frequency is lower than
$2000cm^{-1}$, or $J_2<0.25J_1$. Even this estimate is substantially smaller
than $J_2\sim 0.55J_1$ extracted from the data of infrared transmission and
reflection measurements in $YBCO$~\cite{Gruninger95}. Given that inelastic
neutron scattering measurements were unable to detect the optical spin wave
branch in antiferromagnetic $YBCO$ up to $60meV$, it was
argued~\cite{Shamoto} that the gap, $2 (J_1 J_2)^{1/2}$,
should be larger than $60meV$.
We therefore
conclude that $J_2$ must be in the energy range $8meV<J_2<30meV$.

\centerline{{\bf Acknowledgments}}
It is our pleasure to thank D. Frenkel,  M.V. Klein, A. Millis and H. Monien
for helpful
conversations. A.C. is an A.P. Sloan fellow. A.P.K. gratefully acknowledges
support by the Deutsche Forschungsgemeinschaft through a Heisenberg grant and
the Sonderforschungsbereich 341. G.B. was
 supported by NSF cooperative agreement DMR 91-20000 through the
Science and Technology Center for Superconductivity.

\appendix
\section{Derivation of the Energy Dispersions of the Quasiparticle
Conduction and Valence Bands}
\label{en-disp-app}
In this Appendix we derive the dispersions for the valence and conduction
fermions in the one--band double--layer Hubbard model at half--filling. After
Fourier transformation to momentum space the Hubbard Hamiltonian Eq.
(\ref{Hubbard}) takes the form
\begin{eqnarray}
H&=&\sum_{\bf k,\sigma}{(-4t\nu_{\bf k})(c_{\bf k,\sigma}^\dagger c_{\bf k,
\sigma}+d_{\bf k,\sigma}^\dagger d_{\bf k,\sigma})}-t'\sum_{\bf k}{(c_{\bf k,
\sigma}^\dagger d_{\bf k,\sigma}+d_{\bf k,\sigma}^\dagger c_{\bf k,\sigma})}
\nonumber \\
& &+{U\over 2N}\sum_{\bf k,k',q,\sigma}{\Big(c^\dagger_{\bf k'+q,\sigma}
c^\dagger_{\bf k,-\sigma}c_{\bf -k',-\sigma}c_{\bf k+q,\sigma}+
d^\dagger_{\bf k'+q,\sigma}d^\dagger_{\bf k,-\sigma}d_{\bf -k',-\sigma}
d_{\bf k+q,\sigma}\Big)}\,\, .
\label{Hubbard2}
\end{eqnarray}
The presence of long range antiferromagnetic SDW order in the ground state
implies that
\begin{eqnarray}
{1\over N}\sum_{\bf k}{\langle c^\dagger_{\bf k+\pi,\uparrow}c_{\bf k,\uparrow}
\rangle}&=&-{1\over N}\sum_{\bf k}{\langle c^\dagger_{\bf k+\pi,\downarrow}
c_{\bf k,\downarrow}\rangle=m\neq 0}\, ,  \nonumber \\
{1\over N}\sum_{\bf k}{\langle d^\dagger_{\bf k+\pi,\downarrow}d_{\bf k,
\downarrow}\rangle}&=&-{1\over N}\sum_{\bf k}{\langle d^\dagger_{\bf k+\pi,
\uparrow} d_{\bf k,\uparrow}\rangle=m}\,\, .
\label{lro}
\end{eqnarray}
Introducing the linear combinations
\begin{equation}
\alpha_{\bf k,\sigma}={1\over\sqrt{2}}(c_{\bf k\sigma}+d_{\bf k\sigma})\, ,
\qquad\beta_{\bf k\sigma}={1\over\sqrt{2}}(c_{\bf k\sigma}-d_{\bf k\sigma})
\label{rotation}
\end{equation}
and decoupling the interaction term with the expectation values of (\ref{lro}),
the Hubbard Hamiltonian (\ref{Hubbard2}) turns into
\begin{eqnarray}
H&=&{1\over N}{\sum_{\bf k,\sigma}}^{\prime}\Bigg\{ \ (-4t\nu_{\bf k}-t')
(\alpha^\dagger_{\bf k,\sigma}\alpha_{\bf k,\sigma}-\beta^\dagger_{\bf k+\pi,
\sigma}\beta_{\bf k+\pi,\sigma})-U\, m\, {\rm sgn}(\sigma)\Big[
\alpha^\dagger_{\bf k,\sigma}\beta_{\bf k+\pi,\sigma}+\beta^\dagger_{\bf k+\pi,
\sigma}\alpha_{\bf k,\sigma}\Big]    \nonumber \\
& &\hspace{1.5 cm}+(4t\nu_{\bf k}-t')(\alpha^\dagger_{\bf k+\pi,\sigma}
\alpha_{\bf k+\pi,\sigma}-\beta^\dagger_{\bf k,\sigma}\beta_{\bf k,\sigma})-U\,
m\, {\rm sgn}(\sigma)\Big[\alpha^\dagger_{\bf k+\pi,\sigma}
\beta_{\bf k,\sigma}+\beta^\dagger_{\bf k,\sigma}\alpha_{\bf k+\pi,\sigma}
\Big]\Bigg\}\,\, ,
\label{Hubbard3}
\end{eqnarray}
where the primed momentum sum is restricted to the magnetic Brillouin zone. Two
separate Bogolyubov transformations applied to the first and the second part of
the Hamiltonian (\ref{Hubbard3}) yield two pairs of conduction and valence
bands with the dispersions $\pm E^+_{\bf k}=\pm\sqrt{(4t\nu_{\bf k}+t')^2+
\Delta^2}$ and $\pm E^-_{\bf k}=\pm\sqrt{(4t\nu_{\bf k}-t')^2+\Delta^2}$. The
self--consistency condition for $\Delta=Um$ requires that
\begin{equation}
\frac{1}{U}=\frac{1}{2N}~{\sum_{\bf k}}^{\prime}\left[\frac{1}{E^+_{\bf k}}+
\frac{1}{E^-_{\bf k}}\right]\,\, .
\label{gapequation}
\end{equation}

\section{Transverse Susceptibilities in a Double--Layer Antiferromagnet}
\label{der-susc}
The dynamic, transverse spin susceptibility is obtained from the time ordered
correlation function
\begin{equation}
\chi^{+-}_{\alpha\beta}({\bf q,q}',t)=i<TS^+_{{\bf q},\alpha}(t)S^-_{-{\bf q}',
\beta}(0)>\,\,
\label{chi2}
\end{equation}
where the indices $\alpha,\beta=1,2$ denote the layer. In terms of fermion
operators the spin raising and lowering operators $S^{\pm}_{{\bf q},\alpha}$
are expressed through
\begin{equation}
{\bf S}_{{\bf q},1}={1\over N}\sum_{\bf k}\sum_{\mu,\nu}c^\dagger_{\bf k+q,\mu}
\,\vec{\sigma}_{\mu,\nu}\, c_{\bf k,\nu}\, ,\qquad{\bf S}_{{\bf q},2}={1\over
N}\sum_{\bf k}\sum_{\mu,\nu}d^\dagger_{\bf k+q,\mu}\,\vec{\sigma}_{\mu,\nu}\,
d_{\bf k,\nu}
\label{SDW-trans}
\end{equation}
where $\vec{\sigma}$ are the Pauli matrices and the $c$ and $d$ operators
describe the electrons from layer $1$ and $2$, respectively. Summing the RPA
ladder diagram series for the transverse susceptibilities as described in
detail in the literature for a single--layer system \cite{Sch89,Chu92,Kampf93}
leads to the following results (for S=1/2)
\begin{eqnarray}
\chi^{+-}_{11}({\bf q,q},\omega)&=&-{J_1(1-\nu_{\bf q})\over\omega^2-
\Omega_1^2({\bf q})+i\delta}-{J_1(1-\nu_{\bf q}+J_2/2J_1)\over\omega^2-
\Omega_2^2({\bf q})+i\delta} \nonumber \\
\chi^{+-}_{11}({\bf q,q+Q},\omega)&=& \frac{1}{2}~\left[{\omega\over\omega^2-
\Omega_1^2({\bf q})+i\delta}+{\omega\over\omega^2-\Omega_2^2({\bf q})+i\delta}
\right]\,\, ,
\label{chisamelayer}
\end{eqnarray}
when the spins are from the same layer ($\chi^{+-}_{11}=\chi^{+-}_{22}$), and
\begin{eqnarray}
\chi^{+-}_{12}({\bf q,q},\omega)&=&-{J_1(1-\nu_{\bf q})\over\omega^2-
\Omega_1^2({\bf q})+i\delta}+{J_1(1-\nu_{\bf q}+J_2/2J_1)\over\omega^2-
\Omega_2^2({\bf q})+i\delta} \nonumber \\
\chi^{+-}_{12}({\bf q,q+Q},\omega)&=& \frac{1}{2}~\left[{\omega\over\omega^2-
\Omega_1^2({\bf q})+i\delta}-{\omega\over\omega^2-\Omega_2^2({\bf q})+i\delta}
\right]\,\, ,
\label{chidifferentlayers}
\end{eqnarray}
when the spins are from different layers ($\chi^{+-}_{12}=\chi^{+-}_{21}$). The
poles of the susceptibilities are at $\Omega_1({\bf q})$ and $\Omega_2({\bf q})
\equiv\Omega_1({\bf q+Q})$, as they should. Using these results, one can
construct the effective Hamiltonian for the magnon--fermion interaction Eq.
(\ref{intH}) by the requirement that it reproduces the forms of all these
susceptibilities.

\section{Raman Intensities Without Magnon--Magnon Interaction}
\label{raman-int-app}
In this Appendix we present the closed forms of $I_{A_{1g}}$ and $I_{B_{1g}}$
without final state magnon--magnon interaction. The actual calculations have
been performed to leading order in $1/S$. Here, we present the results for
$S=1/2$. We introduce the short notation $\tilde{\Omega}=\Omega/2J_1$ and
\begin{eqnarray}
a&=&\frac{(16 t^2)^2}{J_1\pi^2\hbar}~\left[\frac{2\Delta}{4\Delta^2-\Omega^2}
{}~\frac{J_2}{4J_1}\right]^2, ~~~~ b=\frac{(16t^2)^2}{J_1\pi^2\hbar}~\left[
\frac{2\Delta}{4\Delta^2-\Omega^2}\right]^2 ~\left(1+\frac{J_2}{4J_1}\right)\,
, \nonumber \\
t_+&=&{1-\nu_+\over 1+\nu_+}\ ,\quad t_-={1+\nu_-\over 1-\nu_-}, \hskip1.cm
g({\tilde\Omega})={1\over{2{\tilde\Omega}}}{1\over\sqrt{(1+{J}_2/4{J}_1)^2-
{\tilde\Omega}^2/4}}\, ,
\label{a}
\end{eqnarray}
where

$\nu_{+,-}=-(J_2/4J_1)\pm\sqrt{(1+{J}_2/4{J}_1))^2-{\tilde\Omega}^2/4}\,\,$.

For $A_{1g}$ scattering geometry we then obtain:\\
i) for $0\leq\tilde{\Omega}\leq 2\sqrt{{J}_2/{J}_1}$
\begin{equation}
I_{A_{1g}}(\Omega)=a \ g({\tilde\Omega}) \ {(1-\nu_+)^2\over 1+\nu_+}K(t_+)
\label{IA1g_1}
\end{equation}
where $K$ is the complete elliptic integral of the first kind;\\
ii) for $2\sqrt{{J}_2/{J}_1}\leq\tilde{\Omega}\leq 2\sqrt{1+{J}_2/2{J}_1}$
\begin{equation}
I_{A_{1g}}(\Omega)=a \ g({\tilde\Omega}) \ \Bigg\{{(1-\nu_+)^2\over 1+\nu_+}
K(t_+)+(1-\nu_-)K(t_-)\Bigg\}\,\, ;
\label{IA1g_2}
\end{equation}
iii) for $2\sqrt{1+{J}_2/2{J}_1}\leq \tilde{\Omega} \leq 2(1+{J}_2/4{J}_1)$
\begin{equation}
I_{A_{1g}}(\Omega)=a \ g({\tilde\Omega}) \ \Bigg\{(1-\nu_+)K(t_+)+(1-\nu_-)
K(t_-)
\Bigg\}\,\, .
\label{IA1g_3}
\end{equation}

\noindent For $B_{1g}$ scattering geometry we obtain\\
i) for $0\leq \tilde{\Omega} \leq 2\sqrt{{J}_2/{J}_1}$
\begin{equation}
I_{B_{1g}}(\Omega)=b \ g({\tilde\Omega}) \ (1+\nu_+)\Big[K(t_+)-E(t_+)\Big]
\,\, ,
\label{IB1g_1}
\end{equation}
where $E$ is the complete elliptic integral of the second kind;\\
ii) for $2\sqrt{{J}_2/{J}_1}\leq \tilde{\Omega} \leq 2\sqrt{1+{J}_2/2{J}_1}$
\begin{equation}
I_{B_{1g}}(\Omega)=b \ g({\tilde\Omega}) \ \Bigg\{(1+\nu_+)\Big[K(t_+)-E(t_+)
\Big]+(1-\nu_-)\Big[K(t_-)-E(t_-)\Big]\Bigg\}\,\, ;
\label{IB1g_2}
\end{equation}
iii) for $2\sqrt{1+{J}_2/2{J}_1}\leq\tilde{\Omega}\leq 2(1+{J}_2/4{J}_1)$
\begin{equation}
I_{B_{1g}}(\Omega)=b \ g({\tilde\Omega}) \ \Bigg\{(1-\nu_+)\Big[K(t_+)-E(t_+)
\Big]+(1-\nu_-)\Big[K(t_-)-E(t_-)\Big]\Bigg\}\,\, .
\label{IB1g_3}
\end{equation}

\section{Raman Intensity with Magnon--Magnon Interaction}
\label{magn-int}
In this Appendix we outline our calculations of the full Raman intensity with
final state interaction. Considering repeated two--magnon scattering we sum the
corresponding series of ladder diagrams (see, e.g. \cite{Can92}). The resulting
integral equation for the full vertex function reduces to a set of algebraic
equations which allows for an explicit solution. We skip the details and list
here only the results.

For the Raman intensity in $A_{1g}$ geometry we obtain
\begin{equation}
I_{A_{1g}}(\Omega)\propto{\rm Im}{\alpha(\Omega)\over 1+\displaystyle{\gamma
\over 2}\alpha(\Omega)}
\label{IA1gmm}
\end{equation}
where
\begin{equation}
\gamma=J_2(4J_1)^2\Big(1+{J_2\over 4J_1}\Big), \quad \alpha(\Omega)=P^{(2)}
(\Omega)+2J_1\Big[R^{(1)}(\Omega)E(\Omega)+{J_2\over 4J_1}P^{(1)}(\Omega)A
(\Omega)\Big]\,\, .
\label{gammaalpha}
\end{equation}
Here we defined
\begin{eqnarray}
A(\Omega)&=&{-P^{(1)}(\Omega)+R^{(1)}(\Omega)H(\Omega)\over 1+{1\over 2}J_2
\Big[P^{(0)}(\Omega)-R^{(0)}(\Omega)H(\Omega)\Big]}  \\[.4cm]
H(\Omega)&=&{2J_1 R^{(0)}(\Omega)\over 1+2J_1Q(\Omega)} \\[.4cm]
E(\Omega)&=&-{{1\over 2}J_2 R^{(0)}(\Omega)A(\Omega)+R^{(1)}(\Omega)\over 1+2
J_1 Q(\Omega)} \,\, ,
\end{eqnarray}
and also
\begin{eqnarray}
P^{(m)}(\Omega)&=&{i\over N}\sum_{\bf q}\int{d\omega\over 2\pi}\Bigg({1-\nu_{
\bf q}\over\Omega_1({\bf q})}\Bigg)^m\ G_0({\bf q},\Omega+\omega)\; G_{0}(-{
\bf q},-\omega) \nonumber \\
&=&{1\over N}\sum_{\bf q}\Bigg({1-\nu_{\bf q}\over\Omega_1({\bf q})}\Bigg)^m{1
\over\Omega-2\Omega_1({\bf q})+i\delta}
\label{P} \\[0.4cm]
R^{(m)}(\Omega)&=&{i\over N}\sum_{\bf q}\int{d\omega\over 2\pi}\Bigg({1-\nu_{
\bf q}\over\Omega_1({\bf q})}\Bigg)^m\; \nu_{\bf q}\ G_{0}({\bf q},\Omega+
\omega) \; G_{0}(-{\bf q},-\omega) \nonumber \\
&=&{1\over N}\sum_{\bf q}\Bigg({1-\nu_{\bf q}\over\Omega_1({\bf q})}\Bigg)^m
{\nu_{\bf q}\over\Omega-2\Omega_1({\bf q})+i\delta}
\label{R} \\[0.4cm]
Q(\Omega)&=&{i\over N}\sum_{\bf q}\int{d\omega\over 2\pi}\; \nu_{\bf q}^2 \
G_{0}({\bf q},\Omega+\omega) \; G_{0}(-{\bf q},-\omega)  \nonumber \\
&=&{1\over N}\sum_{\bf q}{\nu_{\bf q}^2\over\Omega-2\Omega_1({\bf q})+i\delta}
\,\, .
\label{Q}
\end{eqnarray}

where $G_{0}({\bf q},\omega)=(\omega-\Omega_1({\bf q})+i\delta)^{-1}$ is the
non--interacting spin wave propagator.

For $B_{1g}$ geometry we find
\begin{equation}
I_{B_{1g}}(\Omega)\propto{\rm Im}{L^{(2)}(\Omega)-4J_1\Big[L^{(1)}(\Omega)
L^{(1)}(\Omega)-L^{(0)}(\Omega)L^{(2)}(\Omega)\Big]\over 1+4J_1L^{(0)}(\Omega)+
16J_1^2(J_2+4J_1)^2\Big[L^{(0)}(\Omega)L^{(2)}(\Omega)-L^{(1)}(\Omega)L^{(1)}
(\Omega)+\displaystyle{L^{(2)}(\Omega)\over 4J_1}\Big]}
\end{equation}
where
\begin{eqnarray}
L^{(m)}(\Omega)&=&{i\over N}\sum_{\bf q}\int{{d\omega\over 2\pi}\ {\tilde{\nu}_
{\bf q}^2\over\Omega_1^m({\bf q})}\ G_{0}({\bf q},\Omega+\omega)\ G_{0}
(-{\bf q},-\omega)} \nonumber \\
&=&{1\over N}\sum_{\bf q}\ {\tilde{\nu}_{\bf q}^2\over\Omega_1^m({\bf q})}\
{1\over\Omega-2\Omega_1({\bf q})+i\delta}\,\, .
\end{eqnarray}
This form is similar to the result for a single layer. The plots of the full
intensities are presented in Figs.~\ref{B1g-int-fig1} and \ref{A1g-int-fig}.

\begin{figure}
\caption{The system under consideration is a two layer antiferromagnet with
intralayer exchange coupling $J_1=4t^2/U$ and interlayer exchange coupling
$J_2=4t^{' \, 2}/U$.}
\label{system}
\end{figure}

\begin{figure}
\caption{The diagrams which contribute to lowest order in $t/U$ to the Raman
matrix element $M_R$. The fermions from the valence (conduction) bands are
denoted by a dashed (solid) line. The emitted magnons are represented by a
solid wavy line whereas the incoming ($\omega_i$) and outgoing ($\omega_f$)
photons are given by dashed wavy lines.}
\label{non-res-diag}
\end{figure}

\begin{figure}
\caption{The Raman intensities in $A_{1g}$ and $B_{1g}$ geometry obtained in
the Loudon--Fleury theory neglecting final state magnon--magnon interactions.
The jump in the intensity in the $A_{1g}$ geometry occurs at the frequency
$2\Omega_1({\bf Q})=4(J_2J_1)^{1/2}$. The overall shape of the intensities is
shifted towards higher frequencies with increasing $J_2/J_1$.}
\label{B1g-intens}
\end{figure}

\begin{figure}
\caption{The Loudon--Fleury Raman intensity in $B_{1g}$ geometry with final
state interaction for two different values of $J_2/J_1$.}
\label{B1g-int-fig1}
\end{figure}

\begin{figure}
\caption{
The Loudon--Fleury Raman intensity in $A_{1g}$ geometry with final state
interaction. The figures (a), for $J_2/J_1=0.2$, and (b), for $J_2/J_1=0.4$,
include effects of magnon damping which were modeled
by adding a finite imaginary part $i\delta$ to the
energy denominators of the spin wave propagators.
The transferred
frequency in these two figures
is given in units of the maximum spin wave frequency
$\Omega_{max}$.
The low frequency peak in the intensity is located at a frequency somewhat
smaller than $2\Omega_1({\bf Q})$. With increasing damping the
peak frequency is gradually shifted closer to $2\Omega_1({\bf Q})$.
For comparison, in Fig. (c) we plotted the intensity for $J_2/J_1 =0.4$ without
any magnon damping.}
\label{A1g-int-fig}
\end{figure}

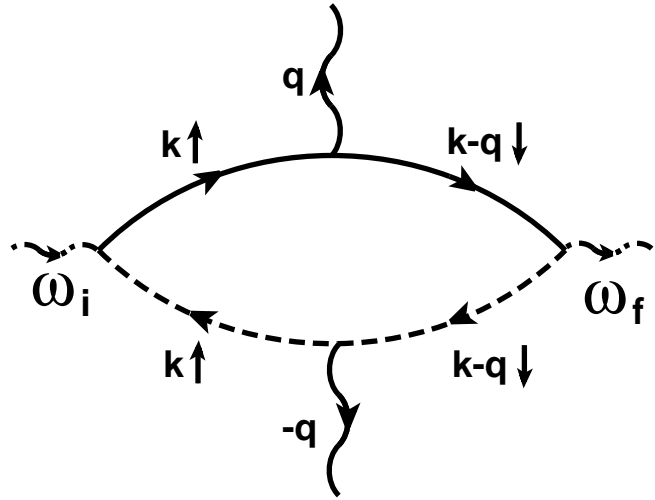
\begin{figure}
\caption{The ``triple--resonance'' diagram which gives the dominant
contribution to the Raman intensity in the resonant regime. The notations are
the same as in Fig.\protect\ref{non-res-diag}}
\label{res-diag}
\end{figure}

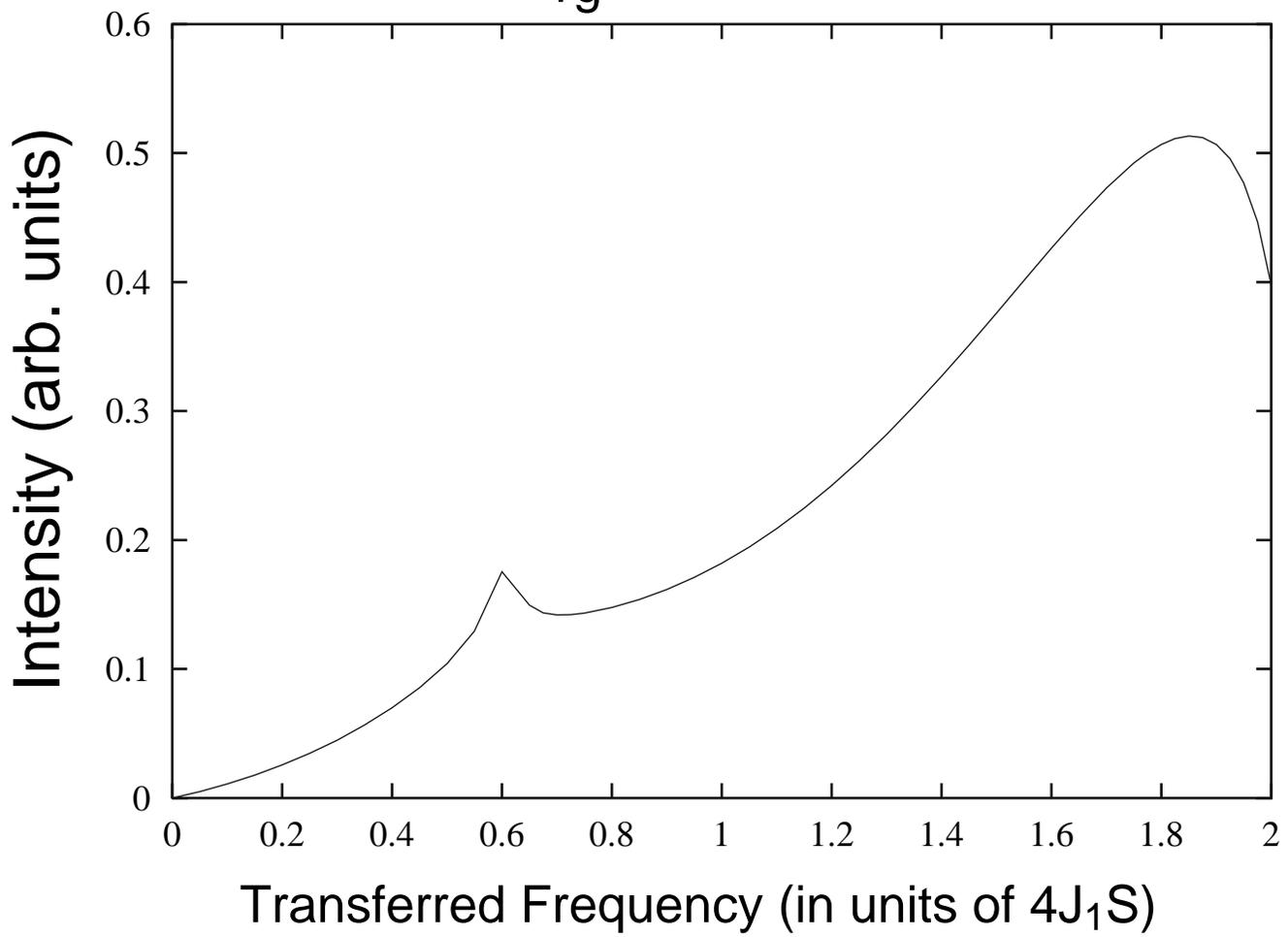
\begin{figure}
\caption{The ``triple--resonance'' diagram contribution to the Raman intensity
in $A_{1g}$ geometry. The final state interaction is included, and
$J_2/J_1=0.1$. Observe the flattening of the intensity above the threshold.
There should be (in the absence of damping) a real jump at the threshold
frequency -- its smearing in the figure is due to limited numerical accuracy.
The dashed line represents the intensity in the absence of $J_2$.}
\label{triple_fig}
\end{figure}

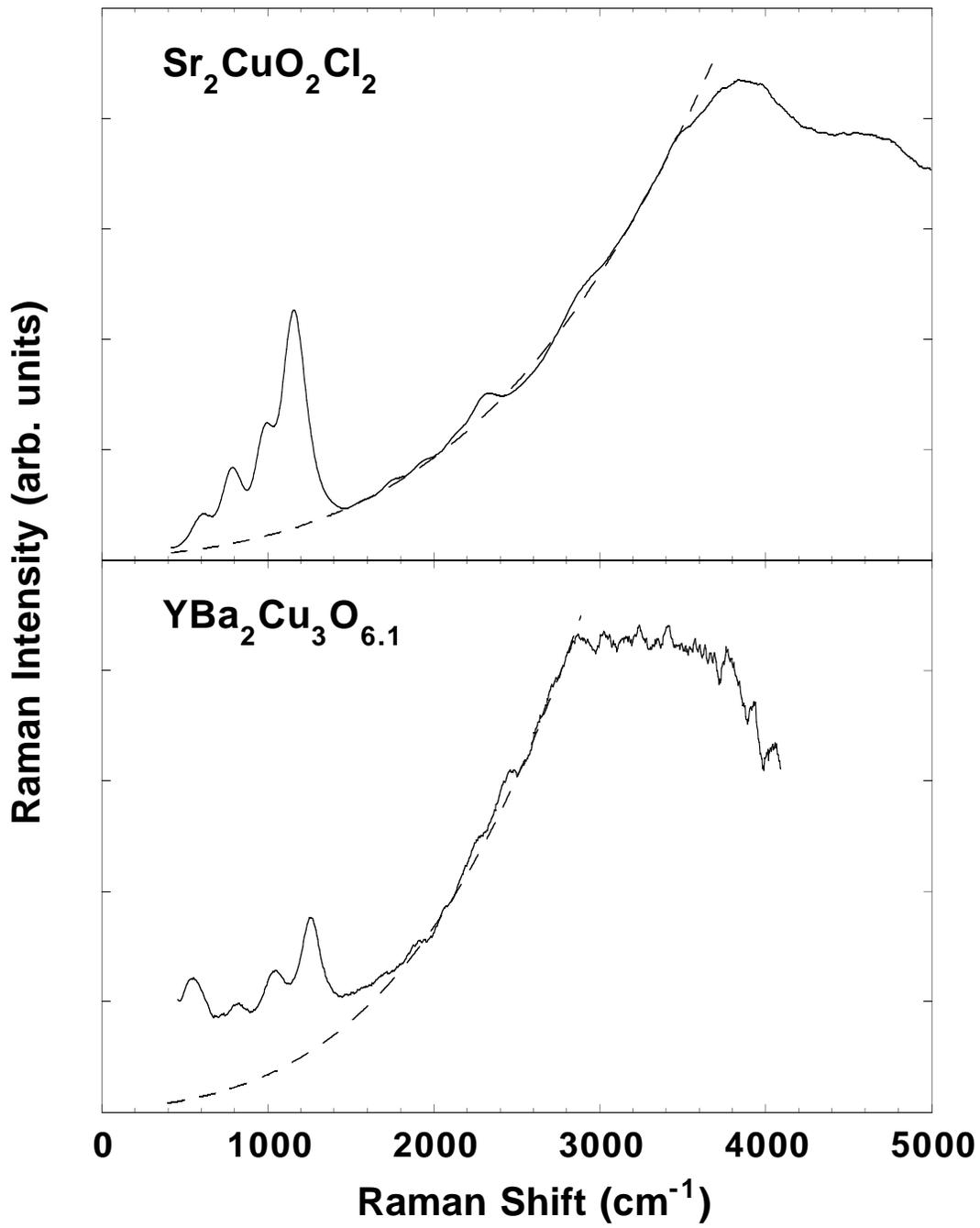
\begin{figure}
\caption{
Experimental Raman scattering
data in $x^{\prime}x^{\prime}$ scattering geometry
 for (a) single-layer $Sr_2CuO_2Cl_2$ and (b)
double-layer $YBa_2Cu_3O_{6.1}$ single crystals. The data are taken from
Ref.\protect\cite{Blumberg95}.
 Observe the flattening of the intensity in $YBCO$ at around $1800cm^{-1}$.
For both crystals, the excitation energy $\omega_{i}$ is
$2 \Delta + 2.9J_{1}$ (the actual values are $2.33$ and 2.09~eV
correspondingly).
The continuum intensity at high frequencies
is presumably due to multi-magnon Raman scattering, the sharp
peaks at low energies are due to resonant multi-phonon
scattering that becomes strongly enhanced for excitations close to
$2\Delta$. The dashed line is a fit
to  linear + cubic frequency dependence
$I \propto [c (\omega/J_{1}) + (\omega/J_{1})^{3}]$, where $c = 1.6$ for
$Sr_2CuO_2Cl_2$ and 1.3 for $YBa_2Cu_3O_{6.1}$.}
\label{exp}
\end{figure}

\begin{references}
\bibitem{xy} Strictly speaking, this scattering geometry corresponds to
$B_{1g}+A_{2g}$, however, we obtained in our theoretical analysis that the
$A_{2g}$ scattering is zero. Experimentally it was found that the signal in
$A_{2g}$ channel is very weak.
\bibitem{xx} This scattering geometry actually yields scattering in the
$A_{1g}$ and $B_{2g}$ channels. However, the $B_{2g}$ intensity is relatively
weak and we will assume that it can be neglected compared to the $A_{1g}$
intensity.
\bibitem{co} We do not consider a case when the light is polarized along the
direction perpendicular to the planes.
\end{references}
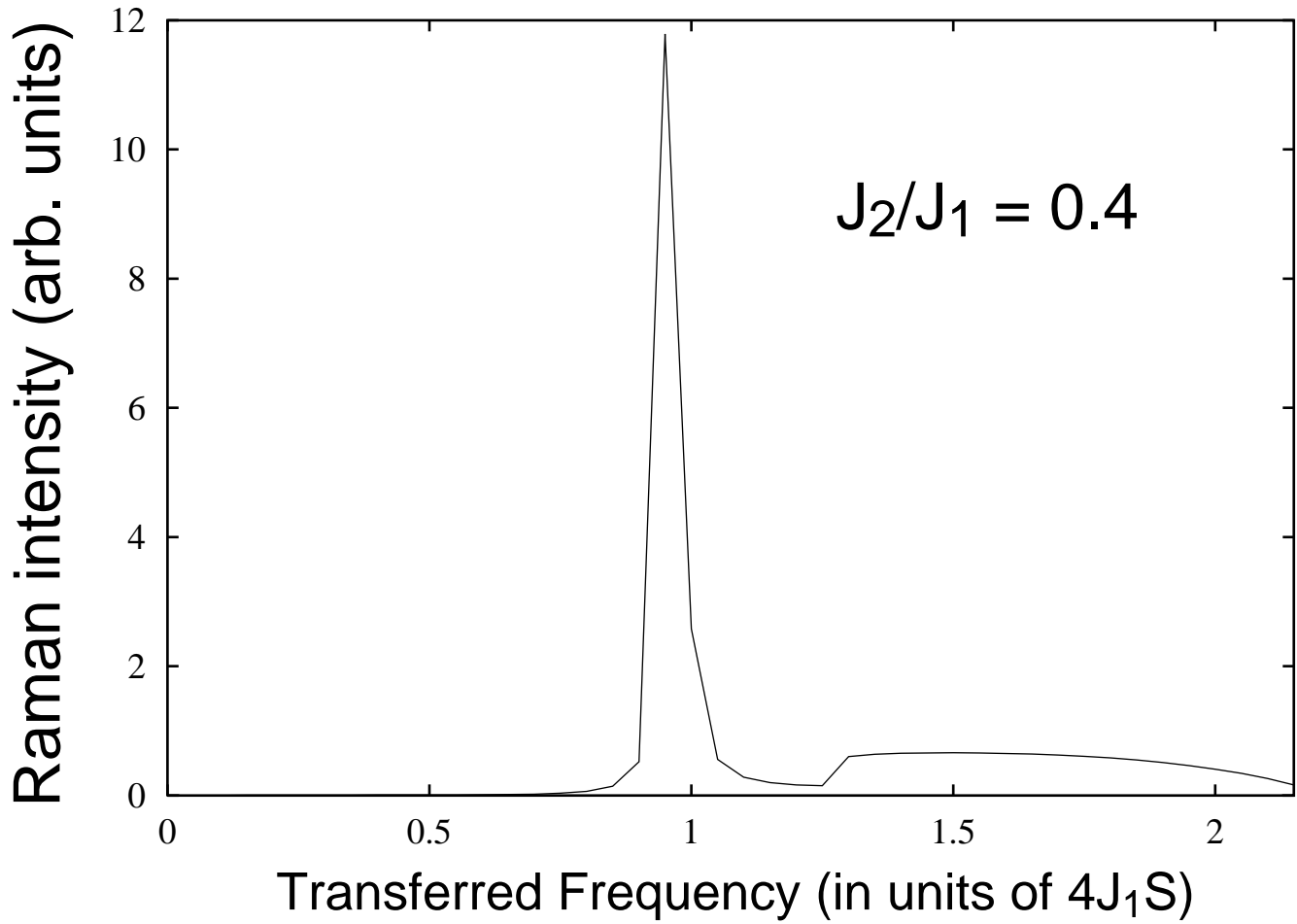
\end{document}